\documentclass{jfm}

\usepackage{amsmath,amssymb,amsfonts,amscd,latexsym,multirow}
\usepackage[utf8]{inputenc}
\usepackage[colorlinks]{hyperref}

\usepackage{graphicx}
\usepackage[export]{adjustbox}
\usepackage{newtxtext}
\usepackage{newtxmath}
\usepackage{natbib}
\usepackage{hyperref}
\hypersetup{
    colorlinks = true,
    urlcolor   = blue,
    citecolor  = black,
}

\newcommand{\RomanNumeralCaps}[1]
\linenumbers

\renewcommand{\leq}{\leqslant}

\renewcommand{\i}{\mathrm{i}}

\newcommand{\am}{\mathrm{am}}

\newcommand{\dn}{\mathrm{dn}}
\newcommand{\K}{\mathrm{K}}
\newcommand{\sech}{\mathrm{sech}}

\title{Almost extreme waves}

\author{Sergey~A.~Dyachenko\aff{1},
\corresp{\email{sergeydy@buffalo.edu}}
Vera~Mikyoung~Hur\aff{2}
\corresp{\email{verahur@math.uiuc.edu}}
\and Denis~A.~Silantyev\aff{3}
\corresp{\email{dsilanty@uccs.edu}} 
}

\affiliation{\aff{1}Department of Mathematics, University of Buffalo, Buffalo, NY 14260-2900 USA
\aff{2}Department of Mathematics, University of Illinois at Urbana-Champaign, Urbana, IL 61801 USA
\aff{3}Department of Mathematics, University of Colorado Colorado Springs, Colorado Springs, CO 80918 USA
}

\begin{document}
\maketitle

\begin{abstract}
Numerically computed with high accuracy are periodic traveling waves at the free surface of a two dimensional, infinitely deep, and constant vorticity flow of an incompressible inviscid fluid, under gravity, without the effects of surface tension. Of particular interest is the angle the fluid surface of an almost extreme wave makes with the horizontal. Numerically found are: (i) a boundary layer where the angle rises sharply from $0^\circ$ at the crest to a local maximum, which converges to $30.3787\dots^\circ$ as the amplitude increases toward that of the extreme wave, independently of the vorticity, (ii) an outer region where the angle descends to $0^\circ$ at the trough for negative vorticity, while it rises to a maximum, greater than $30^\circ$, and then falls sharply to $0^\circ$ at the trough for large positive vorticity, and (iii) a transition region where the angle oscillates about $30^\circ$, resembling the Gibbs phenomenon. Numerical evidence suggests that the amplitude and frequency of the oscillations become independent of the vorticity as the wave profile approaches the extreme form. 
\end{abstract}

\begin{keywords}\end{keywords}


\section{Introduction}\label{sec:intro}

\cite{Stokes1847, Stokes1880} made significant contributions to periodic traveling waves at the free surface of an incompressible inviscid fluid in two dimensions, under gravity, without the effects of surface tension. Particularly, he observed that crests become sharper and troughs flatter as the amplitude increases, and the so-called extreme wave or wave of greatest height displays a $120^\circ$ corner at the crest. Such extreme wave bears relevance to breaking, whitecapping and other physical scenarios. When the flow is irrotational (zero vorticity), based on the reformulation of the problem via conformal mapping as Babenko's nonlinear pseudo-differential equation (see \eqref{eqn:babenko0}), impressive progress was achieved analytically \citep[see, for instance,][]{BDT2000a,BDT2000b} and numerically \citep[see, for instance,][]{DLK2013,DLK2016,Lushnikov2016,LDS2017}.

For zero vorticity, the angle the fluid surface of the extreme wave makes with the horizontal $=30^\circ$ at the crest and $<30^\circ$ at least near the crest \citep[see, for instance,][]{AF1987,McLeod1987}. \cite{Krasovskii1960, Krasovskii1961} conjectured that the angle of any Stokes wave $\leq 30^\circ$. So it came as a surprise when \cite{LHF1} gave analytical and numerical evidence that the angle of an `almost' extreme wave can exceed $30^\circ$ by about $0.37^\circ$ near the crest. 
\cite{LHF2} took matters further and discovered that the wave speed and several other quantities are not monotone functions of the amplitude but, instead, have maxima and minima within a range of the parameter. \cite{McLeod} ultimately proved that Krasovski\u{i}'s conjecture is false. \cite{CG1993} numerically solved Nekrasov's nonlinear integral equation (see \eqref{eqn:nekrasov}) to find that: the angle of an almost extreme wave rises sharply from $0^\circ$ at the crest to approximately $30.3787^\circ$ in a thin boundary layer, oscillates about $30^\circ$, resembling the Gibbs phenomenon, and the angle falls to $0^\circ$ at the trough after the oscillations die out (see also Figure~\ref{fig:w=0}).

Most of the existing mathematical treatment of Stokes waves assume that the flow is irrotational, so that the stream function is harmonic inside the fluid. On the other hand, vorticity has profound effects in many circumstances, for instance, for wind waves or waves in a shear flow. Stokes waves in rotational flows have had a major renewal of interest during the past two decades. We refer the interested reader to, for instance, \citep{vorticity-survey} and references therein. Constant vorticity is of particular interest because one can adapt the approaches for zero vorticity. Also it is representative of a wide range of physical scenarios \citep[see, for instance,][for more discussion]{PTdS1988}. 

For large values of positive constant vorticity, \cite{SS1985} \citep[see also][for finite depth]{PTdS1988} numerically found overhanging profiles and, taking matters further, profiles which intersect themselves tangentially above the trough to enclose a bubble of air. For zero vorticity, by contrast, the wave profile must be the graph of a single-valued function. Here we distinguish positive vorticity for upstream propagating waves and negative vorticity for downstream \citep[see, for instance,][for more discussion]{PTdS1988}. Recently, \cite{DH1,DH2} \citep[see also][]{DH3} offered persuasive numerical evidence that for any constant vorticity, Stokes waves are ultimately limited by an extreme wave in the (amplitude) $\times$ (wave speed) plane, which in the zero vorticity case displays a $120^\circ$ corner at the crest. See Appendix~\ref{appn:angle} for analytical evidence, similar to \citep{Stokes1847,Stokes1880}, that for any constant vorticity, an extreme wave displays a $120^\circ$ corner at the crest.

Here we numerically solve the Babenko equation, modified to accommodate the effects of constant vorticity (see \eqref{eqn:babenko}), with unprecedentedly high accuracy, to discover boundary layer and the Gibbs phenomenon near the crest, alongside other properties of almost extreme waves, in meticulous detail. 
We offer persuasive numerical evidence that for any constant vorticity, the wave speed oscillates as the amplitude increases monotonically toward that of the extreme wave (see Figure~\ref{fig:c(s)}). We predict that 
\[
\frac{c_\text{ext}-c}{\sqrt{s_\text{ext}-s}^3}=\alpha\cos\Big(\tfrac{3}{\pi}K\log(\beta(s_\text{ext}-s))+\gamma\Big)+\cdots\quad \text{as $s\to s_\text{ext}$}
\]
for some constants $\alpha$, $\beta$ and $\gamma$, depending on the vorticity, where $c$ denotes the wave speed, $s$ the steepness, the crest-to-trough vertical distance divided by the period, $c_\text{ext}$ and $s_\text{ext}$ for the extreme wave, and $K=1.1220\dots$ is a positive real root of 
\[
K\tanh K=\frac{\pi}{2\sqrt{3}},
\]
independently of the vorticity. For zero vorticity, see, for instance, \citep{LHF1,LHF2} for more discussion. Also we numerically find that:
\begin{itemize}
\item for any constant vorticity, there is a boundary layer where the angle the fluid surface of an almost extreme wave makes with the horizontal rises sharply from $0^\circ$ at the crest to a (first) local maximum, which converges monotonically to $30.3787\dots^\circ$ as the steepness increases toward that of the extreme wave, independently of the vorticity; the thickness of the boundary layer $\propto (s_{\rm ext}-s)$ as $s\to s_\text{ext}$;
\item there is an outer region where the angle descends monotonically to $0^\circ$ at the trough for zero and negative constant vorticity, while it rises to a maximum $>30^\circ$ and then falls sharply to $0^\circ$ at the trough for large positive vorticity; and
\item there is a transition region where the angle oscillates about $30^\circ$, resembling the Gibbs phenomenon, and the number of oscillations increases as $s$ increases toward that of the extreme wave; the first local minimum converges monotonically to $29.9953\dots^\circ$ as $s\to s_{\rm ext}$, independently of the vorticity. Numerical evidence suggests that the amplitude and frequency of the angle oscillations reach a limit, independent of the vorticity, as $s\to s_{\rm ext}$.
\end{itemize}
See Figures~\ref{fig:w=0}-\ref{fig6}. 

It is difficult to accurately resolve boundary layer and the Gibbs phenomenon because the boundary layer is thin and the angle decreases about two orders of magnitude from one critical value (maximum or minimum) to the next. We solve the modified Babenko equation efficiently using the Newton-conjugate residual method, with aid of an auxiliary conformal mapping, 
to approximate hundreds decimal digits of the steepness at least up to $s_{\rm ext}-s=O(10^{-18})$. See Section~\ref{sec:method} for details. Our result improves \citep{CG1993}, \citep{DH1,DH2} and others. 

\section{Preliminaries}\label{sec:prelim}

Consider a two dimensional, infinitely deep, and constant vorticity flow of an incompressible inviscid fluid, under gravity, without the effects of surface tension, and waves at the fluid surface. We assume for simplicity the unit fluid density. Suppose for definiteness that in Cartesian coordinates, waves propagate in the $x$ direction and gravity acts in the negative $y$ direction. Suppose that the fluid occupies a region $D(t)$ in the $(x,y)$ plane at time $t$, bounded above by a free surface $S(t)$. 

\begin{subequations}\label{eqn:ww}
Let $\boldsymbol{u}(x,y,t)$ denote the velocity of the fluid at the point $(x,y)$ and time $t$, and $P(x,y,t)$ the pressure. They satisfy the Euler equations for an incompressible fluid,
\begin{equation}\label{eqn:euler}
\boldsymbol{u}_t+(\boldsymbol{u}\cdot\nabla)\boldsymbol{u}=-\nabla P+(0,-g)
\quad\text{and}\quad 
\nabla\cdot\boldsymbol{u}=0\quad\text{in $D(t)$},
\end{equation}
where $g$ is the constant acceleration of gravity. We assume that the vorticity
\begin{equation}\label{eqn:vorticity}
\text{$\omega:=\nabla\times\boldsymbol{u}$ is constant throughout $D(t)$.}
\end{equation}
The kinematic and dynamic boundary conditions are
\begin{equation}\label{eqn:bdry surface}
\text{$\partial_t+\boldsymbol{u}\cdot\nabla$ is tangential to $\bigcup_t S(t)$} 
\quad\text{and}\quad P=P_{\rm atm}\quad\text{at $S(t)$},
\end{equation}
where $P_{\rm atm}$ is the constant atmospheric pressure. 
\end{subequations}

Let
\begin{equation}\label{def:phi}
\boldsymbol{u}(x,y,t)=(-\omega y,0)+\nabla\phi(x,y,t),
\end{equation}
so that 
\[
\nabla^2\phi=0\quad\text{in $D(t)$}
\]
by the latter equation of \eqref{eqn:euler}. Namely, $\phi$ is a velocity potential. We pause to remark that for non-constant vorticity, such a velocity potential is no longer viable to use. Substituting \eqref{def:phi} into the former equation of \eqref{eqn:euler} and recalling the latter equation of \eqref{eqn:bdry surface}, after some algebra we arrive at
\[
\phi_t+\tfrac12|\nabla\phi|^2-\omega y\phi_x+\omega\psi+P-P_{\rm atm}+gy=B(t)\quad\text{in $D(t)$},
\]
where $\psi$ is a harmonic conjugate of $\phi$ and $B$ is an arbitrary function. Since $\phi$ and $\psi$ are defined up to addition by functions of $t$, we may assume without loss of generality that 
\begin{equation}\label{eqn:bdry bottom}
\phi,\psi\rightarrow0\quad\text{as $y\rightarrow-\infty$}\quad\text{uniformly for $x$}
\end{equation} 
for all time.

We restrict the attention to traveling wave solutions to \eqref{eqn:ww} and \eqref{eqn:bdry bottom}. That is, $D$, $\phi$ and $\psi$ are stationary in a frame of reference moving with a constant velocity. Let
\[
D=\{(x(u,v),y(u,v)): u\in\mathbb{R}~\text{and}~v<0\}
\quad\text{and}\quad
S=\{(x(u,0),y(u,0)):u\in\mathbb{R}\},
\]
and \eqref{eqn:ww} and \eqref{eqn:bdry bottom} become
\begin{equation}\label{eqn:stokes}
\left\{\begin{aligned}
&\nabla^2\phi, \nabla^2\psi=0 &&\text{in $D$}, \\
&(\phi_x-\omega y-c)y_u=\phi_yx_u &&\text{at $S$},\\
&\tfrac12(\phi_x+\omega y-c)^2+\tfrac12\phi_y^2+gy=B &&\text{at $S$},\\
&\phi,\psi\to0 &&\text{as $y\to-\infty$}\quad\text{uniformly for $x$},
\end{aligned}\right.
\end{equation}
for some $c\neq0$, the wave speed, where $B$ is an arbitrary constant. After the change of variables 
\[
y\mapsto y+y_0\quad\text{and}\quad c\mapsto c-\omega y_0\quad\text{for some $y_0\in\mathbb{R}$},
\]
we may assume that $B=0$. See, for instance, \citep{DH2} for details. Additionally we assume that $D$ and $\psi$ are periodic in the horizontal direction and symmetric about the vertical lines below the crest and trough. We assume without loss of generality that the period is $2\pi$. 

\subsection{The modified Babenko equation}\label{sec:babenko}

Proceeding as in \citep{DH1,DH2} and others, we reformulate \eqref{eqn:stokes} in `conformal coordinates'. In what follows, we identify $\mathbb{R}^2$ with $\mathbb{C}$ whenever it is convenient to do so. 

Suppose that 
\begin{equation}\label{def:conformal}
(x+\i y)(u+\i v) 
\end{equation}
maps $\mathbb{C}_-:=\{u+\i v\in\mathbb{C}:v<0\}$ conformally to $D$ and that  
\[
(x+\i y)(u+\i v)-(u+\i v)
\]
is $2\pi$ periodic in $u$ and $(x+\i y)(u+\i v)-(u+\i v)\to0$ as $v\to-\infty$ uniformly for $u$. Suppose that \eqref{def:conformal} extends to map $\overline{\mathbb{C}_-}$ continuously to $D\bigcup S$. We recall from the theory of Fourier series \cite[see, for instance,][and references therein]{DH1,DH2} that 
\begin{equation}\label{def:x+iy}
(x+\i y)(u+\i 0)=u+((\mathcal{H}+\i)y)(u+\i0),
\end{equation}
where $\mathcal{H}$ denotes the periodic Hilbert transform, defined as 
\[
\mathcal{H}e^{\i ku}=-\i\,\text{sgn}(k)e^{\i ku},\quad k\in\mathbb{Z}.
\]
Abusing notation, let $(\phi+\i\psi)(u+\i v)=(\phi+\i\psi)((x+\i y)(u+\i v))$ and we recall from the theory of Fourier series that
\begin{equation}\label{def:phi+ipsi}
(\phi+\i\psi)(u+\i0)=((1-\i\mathcal{H})\phi)(u+\i0).
\end{equation}
Substituting \eqref{def:x+iy} and \eqref{def:phi+ipsi} into \eqref{eqn:stokes}, after some algebra we arrive at
\begin{equation}\label{eqn:babenko}
c^2\mathcal{H}y_u-(g+\omega c)y=g(y\mathcal{H}y_u+\mathcal{H}(yy_u))
+\tfrac12\omega^2(y^2+\mathcal{H}(y^2y_u)+y^2\mathcal{H}y_u-2y\mathcal{H}(yy_u)).
\end{equation} 
We refer the interested reader to, for instance, \citep{DH1,DH2} for details. When $\omega=0$ (zero vorticity), \eqref{eqn:babenko} becomes
\begin{equation}\label{eqn:babenko0}
c^2\mathcal{H}y_u-gy=g(y\mathcal{H}y_u+\mathcal{H}(yy_u)),
\end{equation} 
namely the Babenko equation \citep{Babenko1987}. 

A solution of \eqref{eqn:babenko} gives rise to a solution of \eqref{eqn:stokes}, provided that 
\begin{subequations}
\begin{align}
&\text{$u\mapsto u+\mathcal{H}y(u)+\i y(u)$ is injective for all $u\in\mathbb{R}$}\label{eqn:injective}
\intertext{and}
&\text{$(1+\mathcal{H}y_u(u))^2+y_u(u)^2\neq 0$ for all $u\in\mathbb{R}$}.\label{eqn:stagnation}
\end{align}
\end{subequations}
See, for instance, \citep{DH1,DH2} for details. We pause to remark that \eqref{eqn:injective} expresses that the fluid surface does not intersect itself and \eqref{eqn:stagnation} ensures that \eqref{def:conformal} is well-defined throughout $\overline{\mathbb{C}_-}$. \citet{DH1,DH2} offered numerical evidence that the solutions of \eqref{eqn:babenko} can be found even though \eqref{eqn:injective} fails to hold, but such solutions are `physically unrealistic' because the fluid surface intersects itself and the fluid flow becomes multi-valued. Recently, \citet{HW:touching} gave a rigorous proof that there exists a `touching' wave, whose profile intersects itself tangentially at one point above the trough to enclose a bubble of air. 

If \eqref{eqn:stagnation} fails to hold, on the other hand, there would be a stagnation point at the fluid surface, where the velocity of the fluid particle vanishes in the moving frame of reference. Numerical evidence \citep[see, for instance,][]{DH1,DH2} supports that for any constant vorticity, the solutions of \eqref{eqn:babenko} would be ultimately limited by an extreme wave in the (amplitude) $\times$ (wave speed) plane, which would display a corner at the crest. In Appendix~\ref{appn:angle} we give analytical evidence, similar to \citep{Stokes1847,Stokes1880}, that for any value of $\omega$, the angle at the crest would be $120^\circ$. 
Here we are interested in `almost' extreme waves. 

\subsection{The Nekrasov equation}\label{sec:nekrasov}

Let 
\[
\theta=\arctan\left(\frac{y_u}{x_u}\right)
\]
denote the angle the fluid surface makes with the horizontal at the point $(x(u),y(u))$, $u\in[-\pi,\pi]$. When $\omega=0$,
\begin{equation}\label{eqn:nekrasov}
\theta(u)=\frac{1}{3\upi}\int^{\upi}_0
\log\left|\frac{\sin\tfrac12(u+u')}{\sin\tfrac12(u-u')}\right|
\frac{\sin\theta(u')}{\mu+\int^{u'}_0\sin\theta}~du',
\end{equation}
where
\begin{equation}\label{def:mu}
\mu=\frac{1}{3gc}{\textstyle \sqrt{c^2-2gy(0)}^3},
\end{equation}
namely the Nekrasov equation \citep{Nekrasov}.
We refer the interested reader to, for instance, \citep{BDT2000a,BDT2000b} for details. Throughout we use subscripts for partial derivatives and primes for variables of integration. We pause to remark that $\sqrt{c^2-2gy(0)}$ is the speed of the fluid particle at the crest. 

\cite{AFT1982} and others proved that for $\mu=0$ there exists an extreme wave and $\theta(u)\to30^\circ$ as $u\to 0$; \cite{PT2004} proved that $\theta$ decreases monotonically over the interval $u\in[0,\pi]$, so that $\sup_{u\in[0,\upi]}|\theta(u)|=30^\circ$. For $\mu\ll1$, on the other hand, \cite{McLeod} proved that $\sup_{u\in[0,\upi]}|\theta(u)|>30^\circ$. 

For $\mu$ sufficiently small, \cite{CG1993} numerically solved \eqref{eqn:nekrasov} to find that: $\theta$ increases from $\theta(0)=0^\circ$ to a maximum $\approx 30.3787^\circ$ in a boundary layer of the size $O(\mu)$; $\theta$ then oscillates about $30^\circ$, and the number of oscillations increases as $\mu\to0$; and $\theta$ decreases to $\theta(\pi)=0^\circ$ outside of the oscillation region. Here we numerically solve \eqref{eqn:babenko0}, with unprecedentedly high accuracy, to improve the result of \citep{CG1993}, and take matters further to include the effects of constant vorticity. 

\section{Methods}\label{sec:method} 

We write \eqref{eqn:babenko} abstractly as
\[
\mathcal{G}(y,c)=0
\]
and solve it iteratively by means of Newton's method. Let 
\[
y^{(n+1)}=y^{(n)}+\delta y^{(n)},\quad n=0,1,2,\dots,
\]
where $y^{(0)}$ is an initial guess \cite[see, for instance,][for details]{DH1,DH2} and
\begin{equation}\label{eqn:linearized}
\delta\mathcal{G}(y^{(n)},c)\delta y^{(n)}=-\mathcal{G}(y^{(n)},c),
\end{equation}
where $\delta\mathcal{G}(y^{(n)},c)$ linearizes $\mathcal{G}(y,c)$ with respect to $y$ and evaluates $y=y^{(n)}$. We solve \eqref{eqn:linearized} numerically using Krylov subspace methods. We approximate $y^{(n)}$ and $\delta y^{(n)}$ using a discrete cosine transform and compute efficiently using a fast Fourier transform. We treat $\mathcal{H}y^{(n)}$ and others likewise. Once obtaining a convergent solution, we continue it along in $c$. We refer the interested reader, for instance, to \cite{DH1,DH2} for details. 

\subsection{Auxiliary conformal mapping}\label{sec:auxiliary}

In what follows, we employ the notation $z=x+\i y$ and $w=u+\i v$. 

In the $\omega=0$ (zero vorticity) case, \cite{DLK2013, DLK2016} and others gave numerical evidence that an analytic continuation of \eqref{def:conformal} to $\mathbb{C}$ has branch points at $w=2n\pi+\i v_0$, $n\in\mathbb{Z}$, for some $v_0>0$. Also
\[
z(w)-w=\sum_{k\in\mathbb{Z},\leq0} \widehat{z}(k)e^{ikw},\quad
\text{where $|\widehat{z}(k)|\propto \exp(-v_0|k|)$}\quad\text{as $|k|\rightarrow\infty$} 
\]
for $v_0$ sufficiently small. Recall that $v_0\rightarrow 0$ as the wave profile approaches the extreme form. This presents enormous technical challenges for numerical computation. Nevertheless \cite{DLK2016} used $2^{27}(=1.34217728\times 10^{8})$ Fourier coefficients to approximate $32$ decimal digits of the steepness for $v_0=O(10^{-7})$. 

To achieve higher accuracy, \cite{LDS2017} introduced
\begin{equation}\label{def:aux'}
w=2\arctan\big(\varepsilon\tan\tfrac12\zeta\big)\quad\text{and}\quad
\zeta=2\arctan\big(\tfrac{1}{\varepsilon}\tan\tfrac12w\big)
\end{equation}
for some $\varepsilon>0$, to be determined in the course of numerical experiment. Note that \eqref{def:aux'} maps $\mathbb{C}_-$ conformally to $\mathbb{C}_-$, $\mathbb{R}+\i0$ to $\mathbb{R}+\i0$, and \eqref{def:aux'} is $2\pi$ periodic in the real variables. In the $\omega=0$ case, therefore, one may solve (see \eqref{eqn:babenko0})
\begin{equation*}\label{eqn:babenko0'}
c^2\mathcal{H}y_\zeta-gu_\zeta y=g(y\mathcal{H}y_\zeta+\mathcal{H}(yy_\zeta)),\quad \zeta\in\mathbb{R},
\end{equation*}
where $\mathcal{H}$ is the periodic Hilbert transform in the $\zeta$ variable and $u_\zeta$ is the Jacobian of \eqref{def:aux'}. Since $u\approx \varepsilon\zeta$, $\zeta\in\mathbb{R}$, about $\zeta=0$ for $\varepsilon\ll1$, \eqref{def:aux'} maps uniform grid points of $\zeta$ to non-uniform grid points of $u$, concentrating the points about $u=0$. Also the latter equation of \eqref{def:aux'} maps $\i v_0$ to, say, $i\zeta_0=\i v_0/\varepsilon+O((v_0/\varepsilon)^3)$ for $v_0/\varepsilon\ll1$, so that
\[
z(w(\zeta))-w(\zeta)=\sum_{k\in\mathbb{Z},\leq0}\widehat{z}(k)e^{ik\zeta},\quad
\text{where $|\widehat{z}(k)|\propto \exp\big(-\tfrac{v_0}{\varepsilon}|k|\big)$}\quad\text{as $|k|\to\infty$}
\]
for $v_0/\varepsilon\ll1$, provided that there are no singularities of the former equation of \eqref{def:aux'} closer to $\mathbb{C}_-$ than $i\zeta_0$. A straightforward calculation reveals that the former equation of \eqref{def:aux'} has branch points at $\zeta=2n\pi\pm2\arctan(\i/ \varepsilon)=(2n\pm1)\pi\pm2\i\varepsilon+O(\i\varepsilon^3)$, $n\in\mathbb{Z}$, for $\varepsilon\ll1$. One may therefore choose $\varepsilon\approx\sqrt{\tfrac12v_0}$, $v_0\ll1$, so that
\[
z(w(\zeta))-w(\zeta)=\sum_{k\in\mathbb{Z},\leq0}\widehat{z}(k)e^{ik\zeta},\quad\text{where 
$|\widehat{z}(k)|\propto \exp(-\sqrt{2v_0}|k|)$}\quad\text{as $|k|\to\infty$}.
\]
This improves numerical convergence. For instance, \cite{LDS2017} used $O(10^4)$ Fourier coefficients to obtain the same result as what \cite{DLK2013,DLK2016} did with $O(10^8)$ Fourier coefficients.

Here we take matters further and resort to 
\begin{equation}\label{def:aux}
w=2\am\Big(\K(\sqrt{m})\frac{\zeta+\pi}{\pi},\sqrt{m}\Big)-\pi
\end{equation}
for some $m$ in the range $(0,1)$, where $\am$ denotes the Jacobi amplitude; that is, for the elliptic parameter $m$ (rather than the elliptic modulus $k$ such that $m=k^2$), 
\[
\varphi=\am(u,\sqrt{m})=F^{-1}(u,\sqrt{m})\quad\text{and}\quad
u=F(\varphi,\sqrt{m})=\int_0^\varphi\frac{d\varphi'}{\sqrt{1-m\sin^2\varphi}}
\]
is the incomplete elliptic integral of the first kind; 
\[
\K(\sqrt{m})=\int_0^1\frac{dt}{\sqrt{(1-t^2)(1-mt^2)}}
\]
is the complete elliptic integral of the first kind. We refer the interested reader, for instance, to \citep{HaleTee2009} for more discussion. We calculate
\begin{equation}\label{def:q}
\zeta=\frac{\pi}{\K(\sqrt{m})}F\big(\tfrac12w+\pi,\sqrt{m}\big)-\pi.
\end{equation}

Note that \eqref{def:aux} maps $\Big\{\zeta\in\mathbb{C}:-\pi\frac{\K'(\sqrt{m})}{\K(\sqrt{m})}<\text{Im}\,\zeta<0\Big\}$ conformally to $\mathbb{C}_-$, and $\mathbb{R}+\i0$ to $\mathbb{R}+\i0$, where  $\K'(\sqrt{m})=\K(\sqrt{1-m})$, and \eqref{def:aux} and \eqref{def:q} are $2\pi$ periodic in the real variables.  Note that \eqref{def:aux} maps 
$\i \zeta$, $\zeta>0$, to $\i v$, where 
\begin{equation}\label{def:qc}
\zeta=\pi\frac{\K'(\sqrt{m})}{\K(\sqrt{m})}\quad\text{and}\quad 
m=1-\tanh^2\big(\tfrac12v\big)=\sech^2\big(\tfrac12v\big).
\end{equation}
Note that \eqref{def:aux} maps $[0,\i \zeta]$, $\zeta>0$, to $[0,\i v]$, where $\zeta$ and $v$ are in \eqref{def:qc}. Also note that \eqref{def:aux} maps $[-\pi+\i\zeta,0+\i\zeta]$ and $[0+\i\zeta,\pi+\i\zeta]$ to $[\i v,+\i\infty]$, making a branch cut of \eqref{def:aux}. 

A straightforward calculation reveals that 
\[
\zeta_w(\zeta)=\frac{\pi}{2}\frac{\dn\big(\tfrac1\pi\K(\sqrt{m})\zeta,\sqrt{m}\big)}{\sqrt{1-m}\K(\sqrt{m})},
\]
where $\dn$ is a Jacobi elliptic function, defined as $\dn(u,\sqrt{m})=\sqrt{1-m\sin^2(\am(u,\sqrt{m}))}$. Recall that $\dn(\cdot,\sqrt{m})$ has periods $2\K(\sqrt{m})$ and $4\i\K'(\sqrt{m})$, zeros at $(2n+1)\K+(2n'+1)\i\K'$
and simple poles at $2n\K+2n'\i\K'$ for any $n,n'\in\mathbb{Z}$, so that 
\begin{align*}
&\zeta_w(\zeta)=0\quad\;\text{at}\quad\zeta=\pm\pi+(2n+1)\i\pi\frac{\K'(\sqrt{m})}{\K(\sqrt{m})}, \quad n\in\mathbb{Z},
\intertext{and}
&\zeta_w(\zeta)\to\infty\quad\text{at}\quad\zeta=0+2n\i\pi\frac{\K'(\sqrt{m})}{\K(\sqrt{m})},\quad n\in\mathbb{Z}.
\end{align*}

For $v_0>0$ and sufficiently small, where $\i v_0$ the closest singularity of \eqref{def:conformal} to $\mathbb{C}_-$, therefore we choose 
\[
m=1-\tanh^2\big(\tfrac12v_0\big)=\sech^2\big(\tfrac12v_0\big),
\]
so that \eqref{def:aux} maps $[-\pi+\i \zeta_0,0+\i \zeta_0]$ and $[0+\i \zeta_0,\pi+\i \zeta_0]$ to $[\i v_0,+\i\infty]$, that is, the branch cut of \eqref{def:aux} to the branch cut of \eqref{def:conformal}, where 
\[
\zeta_0=\pi\frac{\K'(\sqrt{m})}{\K(\sqrt{m})}\approx \frac{\pi^2}{2}\frac{1}{\log(8/v_0)}
\quad\text{for $v_0\ll1$}.
\]
Correspondingly,
\[
z(w(\zeta))-w(\zeta)=\sum_{k\in\mathbb{Z},\leq0}\widehat{z}(k)e^{ik\zeta},\quad
\text{where }|\widehat{z}(k)|\propto \exp\Big(-\frac{\pi^2}{2}\frac{|k|}{\log(8/v_0)}\Big)\quad\text{as $|k|\to\infty$}
\]
for $v_0\ll1$. This dramatically improves numerical convergence. We use \eqref{def:aux} to approximate almost extreme waves for $v_0$ up to $O(10^{-30})$. 

\subsection{Method for nonzero vorticity: CG vs. CR}\label{sec:nonzero vorticity}

Since 
\begin{align*}
\delta\mathcal{G}(y,c)\delta y=&c^2\mathcal{H}(\delta y)_u-(g+c\omega)\delta y
-g(\delta y\mathcal{H}y_u+y\mathcal{H}(\delta y)_u+\mathcal{H}(y\delta y)_u) \\
&-\tfrac12\omega^2(2y\delta y+\mathcal{H}(y^2\delta y)_u-[2y\delta y,y]+[y^2,\delta y]
\end{align*}
is self-adjoint, where $[f_1,f_2]=f_1\mathcal{H}f_2-f_2\mathcal{H}f_1$, \cite{DH2,DH3} employed the conjugate gradient (CG) method \citep[see, for instance,][]{Yang2010} to numerically solve \eqref{eqn:linearized}. For any value of $\omega$, the CG method converges within a range of the parameters although $\delta\mathcal{G}(y,c)$ is {\em not} positive definite, but the method breaks down as the wave profile approaches the extreme form. Even when the method converges, the solution error is {\em not} a monotonically decreasing function of the number of iterations for almost extreme waves. 

Here we resort to Krylov subspace methods for symmetric indefinite systems, particularly, minimal residual (MINRES) methods. MINRES minimizes the $L^2$-norm of the residual 
and does not suffer from breakdown. See, for instance, \citep{PS1975} for more discussion. Indeed, replacing the CG method by the conjugate residual (CR) method works well for any value of $\omega$ for almost extreme waves, and the solution error is monotonically decreasing. 

We require the truncation error $|\widehat{y^{(n)}}(N/2)|\lesssim 10^{-36}$, where $N$ is the number of Fourier coefficients or, alternatively, the number of uniform grid points in the $\zeta$ variable, and the residual $\|\mathcal{G}(y^{(n)}, c)\|_{L^2}\lesssim 10^{-43}$, to approximate 200 decimal digits of the steepness for $s_{\rm ext}-s$ up to $O(10^{-19})$, where $s$ is the steepness and $s_{\rm ext}$ for the extreme wave.

The wave speed varies exponentially slightly along the solution curve, though, near the extreme wave (see Figure~\ref{fig:c(s)}) and our numerical computation must use arbitrary precision floating point numbers. We use GNU MPFR library for variable precision numbers \citep[see, for instance,][]{GNUMPFR}, increasing the number of bits per floating point number as $s\to s_{\rm ext}$. It turns out that about $2560$ bits suffice for $s_{\rm ext}-s$ up to $O(10^{-19})$.

\subsection{Method for zero vorticity}

For $\omega=0$, alternatively, we solve \eqref{eqn:linearized} non-iteratively because solving a $4096\times 4096$ linear system would suffice to approximate $200$ decimal digits of the steepness for $s_{\rm ext}-s$ up to $O(10^{-18})$. The solution error decreases quadratically, that is, the number of significant digits in the numerical solution increases by a factor of $2$ in each Newton's iteration so long as the method converges. 

\section{Results}\label{sec:result} 

\begin{figure}
\centering
\includegraphics[scale=1,valign=t]{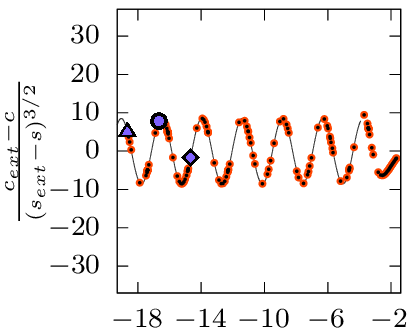}
\includegraphics[scale=1,valign=t]{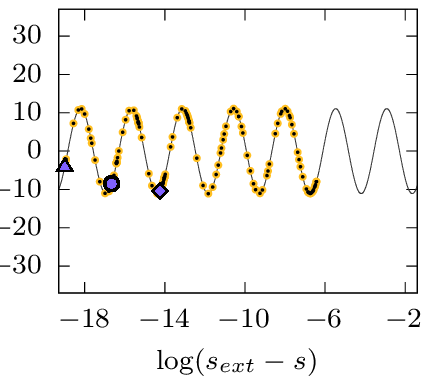}
\includegraphics[scale=1,valign=t]{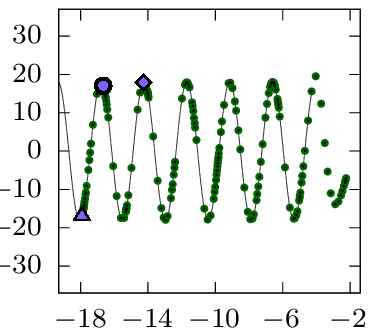}
\caption{
$(c_{\rm ext}-c)/\sqrt{s_{\rm ext}-s}^3$ vs. $\log(s_{\rm ext}-s)$ for $\omega=0$ (left, red), $1$ (middle, yellow), and $-1$ (right, green). Dotted curves are the numerical results and solid for cosine curve fitting. See Figures~\ref{fig:w=0}-\ref{fig:w=-1} for solutions corresponding to the circles, triangles and diamonds.} 
\label{fig:c(s)}
\end{figure}

In the $\omega=0$ (zero vorticity) case, \cite{DLK2016,LDS2017} and others gave numerical evidence that the wave speed converges oscillatorily to that of the extreme wave as the steepness increases monotonically toward that of the extreme wave. The wave speed decreases about two orders of magnitude from one critical value (maximum or minimum) to the next, though, and it is difficult to accurately resolve such wave speed oscillations. Nevertheless \cite{LDS2017} resolved about $3.5$ oscillations, predicting that
\begin{equation}\label{eqn:LH}
\frac{c_{\rm ext}-c}{\sqrt{s_{\rm ext}-s}^3}
=\alpha\cos\Big(\tfrac{\pi}{3}K\log(\beta(s_{\rm ext}-s))+\gamma\Big)+\cdots \quad\text{as $s\to s_{\rm ext}$}
\end{equation}
for some constants $\alpha$, $\beta$ and $\gamma$, where $K=1.1220\dots$ is a positive real root of 
\[
K\tanh K=\frac{\pi}{2\sqrt{3}}.
\]
Here and elsewhere, $c$ denotes the wave speed and $s$ the steepness, the crest-to-trough vertical distance divided by the period, and $c_{\rm ext}$ and $s_{\rm ext}$ for the extreme wave. See also \citep{LHF2} for more discussion. 

The left panel of Figure~\ref{fig:c(s)} shows $(c_{\rm ext}-c)/\sqrt{s_{\rm ext}-s}^3$ versus $\log(s_{\rm ext}-s)$ of our numerical solutions, for $s_{\rm ext}-s$ up to $O(10^{-18})$, and compares the result with \eqref{eqn:LH}, where $c_{\rm ext}$, $s_{\rm ext}$ and $\alpha$, $\beta$, $\gamma$ are determined from the numerics. We exploit an auxiliary conformal mapping (see \eqref{def:aux}) to improve the result of \citep{LDS2017} and others, resolving about $6.5$ oscillations. We report $c_{\rm ext}=1.092285048586537534852655600804383\ldots$ and $s_{\rm ext}=0.14106348397993608209382275\ldots$. 

The middle and right panels of Figure~\ref{fig:c(s)} show $(c_{\rm ext}-c)/\sqrt{s_{\rm ext}-s}^3$ versus $\log(s_{\rm ext}-s)$ for $\omega=1$ and $-1$, for $s_{\rm ext}-s$ up to $O(10^{-19})$, and compare the numerical result with \eqref{eqn:LH}, discovering that $\alpha$, $\beta$ and $\gamma$ depend on $\omega$ but, interestingly, $K$ does {\em not}. We predict that for {\em any} constant vorticity, the wave speed oscillates as $c\to c_{\rm ext}$ while $s\to s_{\rm ext}$ monotonically, and the frequency of the wave speed oscillations is independent of the vorticity. We report $c_{\rm ext}=2.2683602961\dots$ and $s_{\rm ext}=0.4431049878\dots$ for $\omega=1$; $c_{\rm ext}=0.6710639577\dots$ and $s_{\rm ext}=0.0492991750\dots$ for $\omega=-1$.

\begin{figure}
\centerline{
\includegraphics[scale=1]{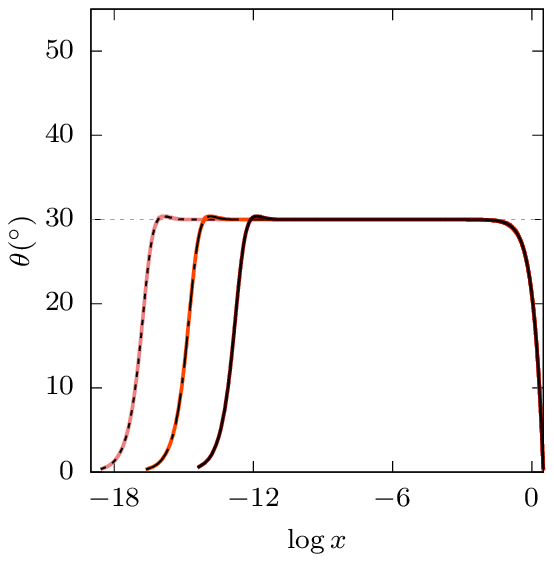}
\includegraphics[scale=1]{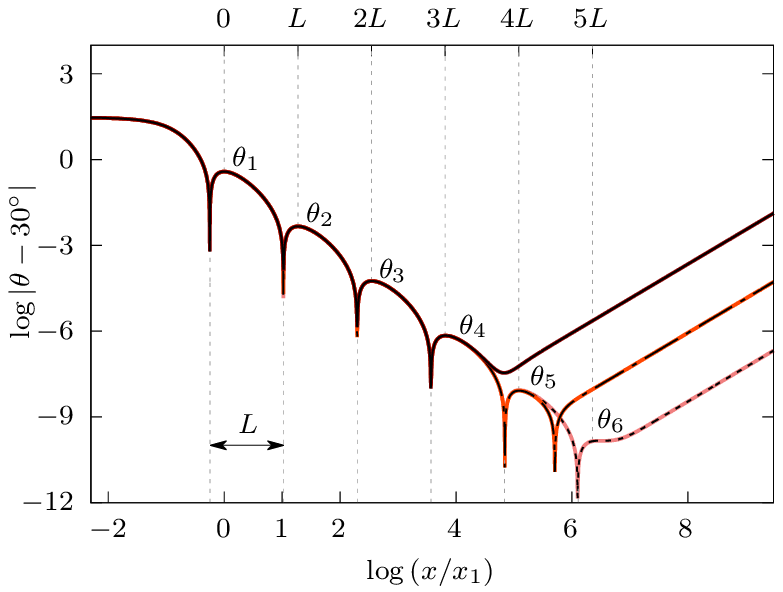}
}
\caption{$\omega=0$. On the left, $\theta (^\circ)$ vs. $\log x$, $x\in[0,\pi]$, for $s_{\rm ext}-s=O(10^{-18})$ (dotted), $O(10^{-16})$ (dashed) and $O(10^{-14})$ (solid). 
On the right, $\log|\theta-30^\circ|$ vs. $\log(x/x_1)$ in the Gibbs oscillation region, where $\theta(x_1)=:\theta_1$ is the first local maximum. See Table~\ref{tab:angle} for approximate values of $\theta_j$, $j=1,2,\dots 6$. Numerically found is $L\approx 2.93$.}
\label{fig:w=0}
\end{figure}

In what follows, by the angle, denoted $\theta (^\circ)$, we mean the angle the fluid surface makes with the horizontal. 

We begin by taking $\omega=0$. The left panel of Figure~\ref{fig:w=0} shows the graph of $\theta$ as a function of $x$, over the interval $x\in[0,\pi]$, for an almost extreme wave for which $s_{\rm ext}-s=O(10^{-18})$, and compares the result with two other almost extreme waves, for which $s_{\rm ext}-s=O(10^{-16})$ and $O(10^{-14})$. Note that the horizontal axis is logarithmic. The three almost extreme waves are marked by the triangle, circle and diamond in the left panel of Figure~\ref{fig:c(s)}. We compute 
$c=1.0922850485865375348526556088099$ (dotted), 
$1.092285048586537534852668$ (dashed) and 
$1.09228504858653753485$ (solid), agreeing on $20$ decimal digits. 

We numerically find a boundary layer where $\theta$ rises sharply from $\theta(0)=0^\circ$ to a (first) local maximum $\theta(x_1)=:\theta_1$, and an outer region where $\theta$ falls to $\theta(\pi)=0^\circ$. We compute $s_{\rm ext}-s=1.3777\ldots\times 10^{-18}$ and $x_1=1.4095\ldots\times 10^{-16}$ (dotted), 
$s_{\rm ext}-s=1.3604\ldots\times 10^{-16}$ and $x_1=1.3918\ldots\times 10^{-14}$ (dashed), $s_{\rm ext}-s=1.3460\ldots\times 10^{-14}$ and $x_1= 1.3771\ldots\times  10^{-12}$ (solid), predicting that $x_1\approx 102.3(s_{\rm ext}-s)$ as $s \to s_{\rm ext}$. Also we numerically find a transition region where $\theta$ oscillates about $30^\circ$, resembling Gibbs phenomenon, although not visible in the scale. Our result agrees with \citep{CG1993} and others. 

The right panel of Figure~\ref{fig:w=0} shows the Gibbs oscillations in the logarithmic scale. We numerically resolve six critical values, denoted $\theta_j:=\theta(x_j)$, $j=1,2,\dots,6$, and  $0<x_1<x_2<\cdots<x_6<\pi$, while observing that the number of oscillations increases as $s$ increases toward $s_{\rm ext}$. Table~\ref{tab:angle} gives approximate critical values for $s_{\rm ext}-s=1.3777\ldots\times10^{-18}$ or, equivalently, $\mu= 2.1978\ldots\times 10^{-26}$ 
(see \eqref{def:mu}) and compares the result with five critical values \cite{CG1993} computed for $\mu=10^{-18}$. Recall that $\mu\to0$ as $s\to s_{\rm ext}$. We predict that 
\begin{equation}\label{eqn:theta12}
\theta_1\to 30.3787032466\dots^\circ\quad\text{and}\quad 
\theta_2 \to 29.9953964674\dots^\circ\quad\text{as $s\to s_{\rm ext}$}.
\end{equation}
Also numerical evidence suggests that 
\begin{equation}\label{eqn:L}
\frac{|\theta_{j+1}-30^\circ|}{|\theta_j-30^\circ|}\approx 1.22\times 10^{-2}
\quad\text{and}\quad
\frac{x_{j+1}}{x_j}\approx 18.73\quad \text{for $s_{\rm ext}-s\ll1$},
\quad j=1,2,\dots,
\end{equation}
or, equivalently, $L:=\log(x_{j+1}/x_1)-\log(x_j/x_1)\approx 2.93$ as $s \to s_{\rm ext}$, independently of $j$. Particularly, $x_j\to0$ as $s\to s_{\rm ext}$ for all $j$.

\begin{figure}
\centerline{
\includegraphics[scale=1]{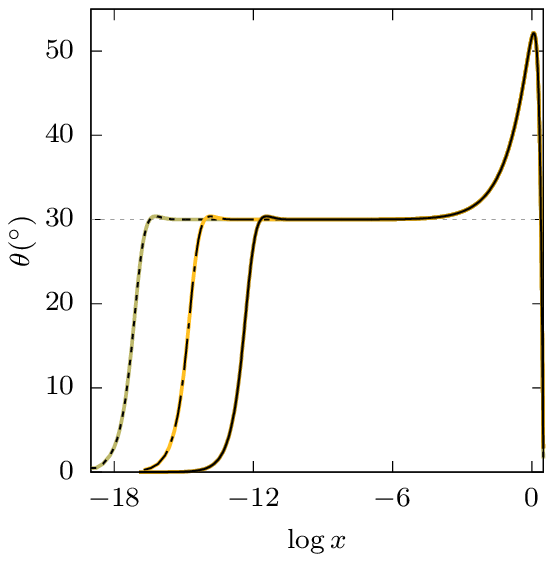}
\includegraphics[scale=1]{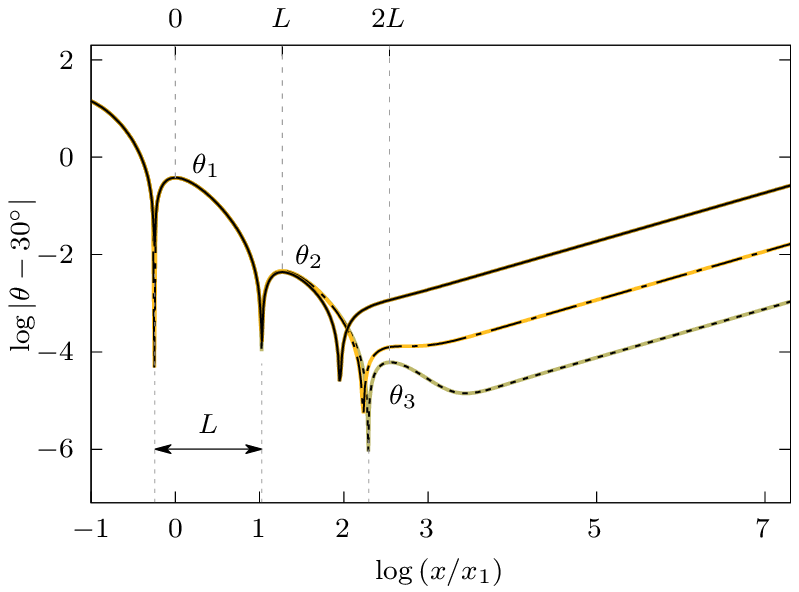}}
\caption{$\omega=1$. (Left) $\theta (^\circ)$ vs. $\log x$, $x\in[0,\pi]$, for $s_{\rm ext}-s=O(10^{-19})$ (dotted), $O(10^{-16})$ (dashed) and $O(10^{-14})$ (solid). (Right) $\log|\theta-30^\circ|$ vs. $\log(x/x_1)$, where $\theta(x_1)$ is the first local maximum. 
See Table~\ref{tab:angle} for approximate values of $\theta_1$, $\theta_2$ and $\theta_3$.
}
\label{fig:w=1}
\end{figure}

\begin{figure}
\centerline{
\includegraphics[scale=1]{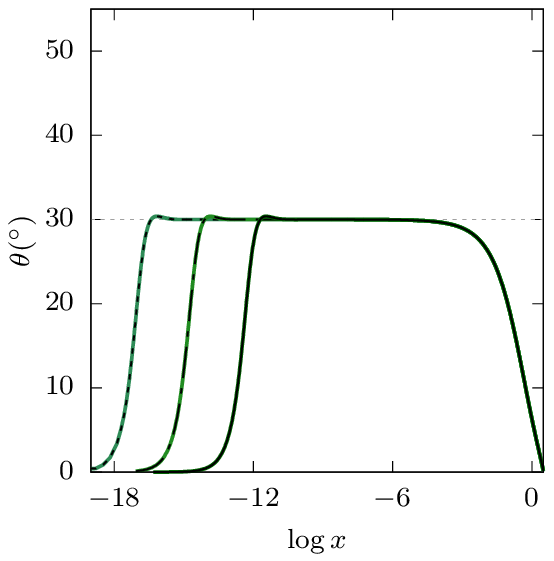}
\includegraphics[scale=1]{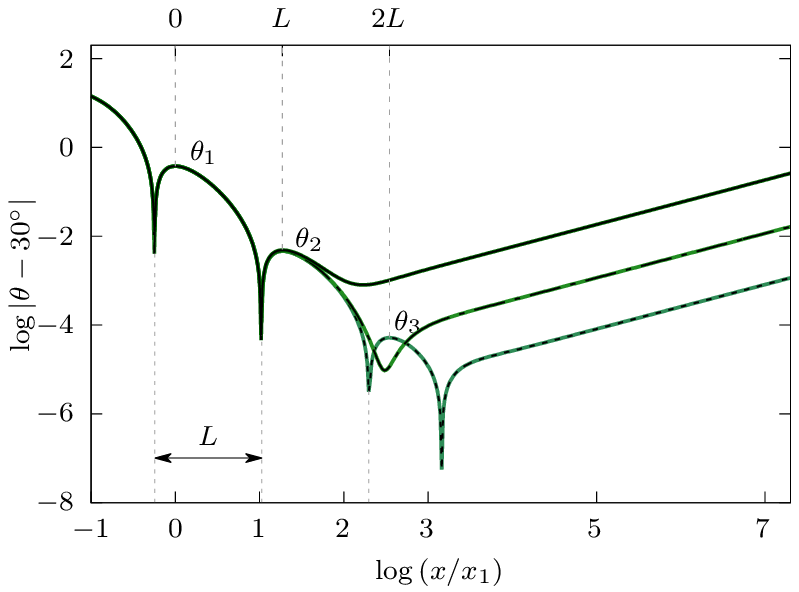}}
\caption{$\omega=-1$. Same as Figures~\ref{fig:w=0} and \ref{fig:w=1}. See Table~\ref{tab:angle} for $\theta_1$, $\theta_2$ and $\theta_3$.}
\label{fig:w=-1}
\end{figure}

We turn the attention to $\omega=1$. Figure~\ref{fig:w=1} shows $\theta$ versus $x$, $x\in[0,\pi]$, in the logarithmic scale, for three almost extreme waves, marked by the triangle, circle and diamond in the middle panel of Figure~\ref{fig:c(s)}. We compute $c= 2.2683602961775079931212358828$ and $s_{\rm ext}-s=O(10^{-19})$ (dotted), $c=2.2683602961775079931212217294$ and $s_{\rm ext}-s=O(10^{-16})$ (dashed), $c=2.2683602961775079930499364611$ and $s_{\rm ext}-s=O(10^{-14})$ (solid). Numerically found are a boundary layer where $\theta$ rises sharply from $\theta(0)=0^\circ$ to a first local maximum $\theta(x_1)=:\theta_1$, where $x_1\approx 102.4(s_{\rm ext}-s)$ as $s \to s_{\rm ext}$, and a Gibbs oscillation region, same as the $\omega=0$ case. We numerically resolve up to the second local maximum angle for $s_{\rm ext}-s=O(10^{-19})$, while observing that higher-order local maxima and minima set in for $s$ closer to $s_{\rm ext}$, compared with the $\omega=0$ case. See Table~\ref{tab:angle} for approximate critical values for $s_{\rm ext}-s=4.8086\dots\times 10^{-19}$. We predict that $\theta_1\to 30.378703256\dots^\circ$ and $\theta_2\to 29.99539\dots^\circ$ as $s\to s_{\rm ext}$, same as the $\omega=0$ case (see \eqref{eqn:theta12}). Also we predict that $L:=\log(x_{j+1}/x_1)-\log(x_j/x_1)\approx 2.93$ as $s\to s_{\rm ext}$, independently of $j$, same as the $\omega=0$ case (see \eqref{eqn:L}). But an important difference is that in the outer region, $\theta$ rises to a maximum $52.1426155193\dots^\circ$ and then falls sharply to $\theta(\pi)=0^\circ$. See the left panel of Figure~\ref{fig:w=1}.

Last but not least, in the $\omega=-1$ case, Figure~\ref{fig:w=-1} shows $\theta$ for $s_{\rm ext}-s=O(10^{-19})$, $O(10^{-16})$ and $O(10^{-14})$, corresponding to the triangle, circle and diamond in the right panel of Figure~\ref{fig:c(s)}. We compute $c = 0.67106395770868248459028861879$, 
$0.67106395770868248459031619969$ and
$0.67106395770868248470517587268$. The result is the same as the $\omega=0$ case, but critical values of the angle set in for $s_{\rm ext}-s$ smaller compared with the $\omega=0$ case.

\begin{table}
\def~{\hphantom{0}}
\begin{tabular}{ c c c c c}
    \hline
        $\omega$ & $s_{\rm ext} - s$ & $\theta_j (^\circ)$ & $|\theta_j-30|(^\circ)$ & \citep{CG1993} \\
        \hline 
        \multirow{6}{*}{$0.0$} & \multirow{6}*{$1.3777\times 10^{-18}$} & 
            $30.3787032466$ & $3.78703246652\texttt{e-01}$ &
            $3.787032466\texttt{e-01}$ \\
        & & $29.9953964674$ & $4.60353262916\texttt{e-03}$ &
            $4.60353\texttt{e-03}$ \\
        & & $30.0000566331$ & $5.66330666364\texttt{e-05}$ &
            $5.6631\texttt{e-05}$ \\
        & & $29.9999993034$ & $6.96605412024\texttt{e-07}$ &
            $7.4218\texttt{e-07}$ \\
        & & $30.0000000086$ & $8.56571838806\texttt{e-09}$ & 
            $3.6722\texttt{e-07}$ \\  
        & & $30.0000000000$ & $1.47128277182\texttt{e-10}$ &\\
        \hline
        \multirow{4}*{$1.0$} 
        & \multirow{4}*{
            $4.8086\times 10^{-19}$} & ${\bf 30.378703246}5$ & $3.78703518762\texttt{e-01}$ & \\
            & & ${\bf 29.99539}74098$                        & $4.60235904573\texttt{e-03}$ & \\
            & & ${\bf 30.00006}07270$                        & $6.17344001164\texttt{e-05}$ & \\
            & & ${\bf 30.0000}127856$                        & $1.56031059113\texttt{e-05}$ & \\
        \hline
        \multirow{3}{*}{$-1.0$} & \multirow{3}*{$6.8836\times 10^{-19}$} &    
            ${\bf 30.378702}9890$ & $3.78702126845\texttt{e-01}$&\\
        & & ${\bf 29.99539}53515$ & $4.60838078515\texttt{e-03}$ &\\
        & & ${\bf 30.00005}18141$ & $3.58968559624\texttt{e-05}$ &\\ \hline
    \end{tabular}
    \caption{Approximate values of $\theta_j$ for $s_{\rm ext}-s\ll1$ for $\omega=0$, $1$ and $-1$, and in the $\omega=0$ case, comparison with \citep{CG1993}. Digits in bold agree up to rounding across numerical computation and also the result for $\omega=0$.} 
    \label{tab:angle}
\end{table}

\begin{figure}
\centering
\includegraphics[scale=1]{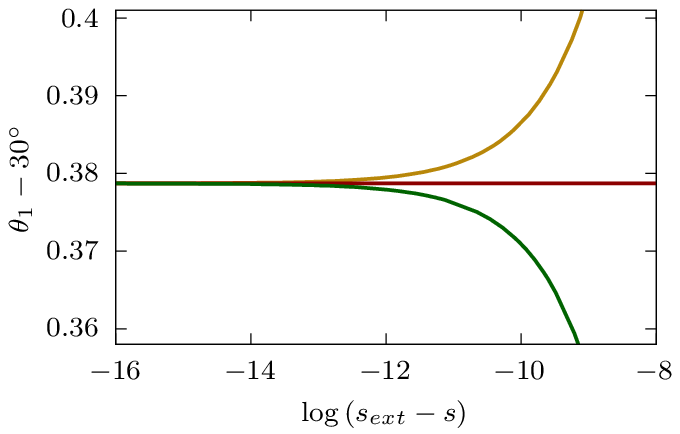}
\includegraphics[scale=1]{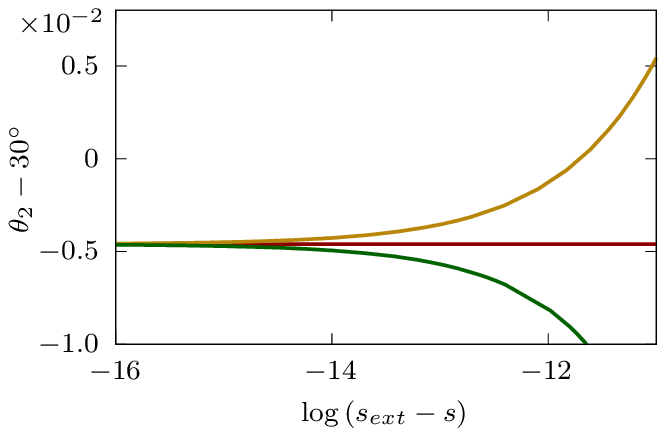}
\caption{$\theta_1$ (left) and $\theta_2$ (right) vs. $\log(s_{\rm ext}-s)$ for $\omega=0$ (red), $1$ (yellow) and $-1$ (green).}
\label{fig:angle}
\end{figure}

Figure~\ref{fig:angle} shows that the first local maximum angle in the oscillation region converges monotonically to $30.3787\dots^\circ$ and the first local minimum to $29.9953\dots^\circ$ as $s\to s_{\rm ext}$, independently of the constant vorticity. We predict that higher-order local maxima and minima converge monotonically as $s\to s_{\rm ext}$, independently of the vorticity. By contrast, the wave speed and several other quantities are {\em not} monotone functions of the steepness (see Figure~\ref{fig:c(s)}).

\begin{figure}
\centering
\includegraphics[scale=1]{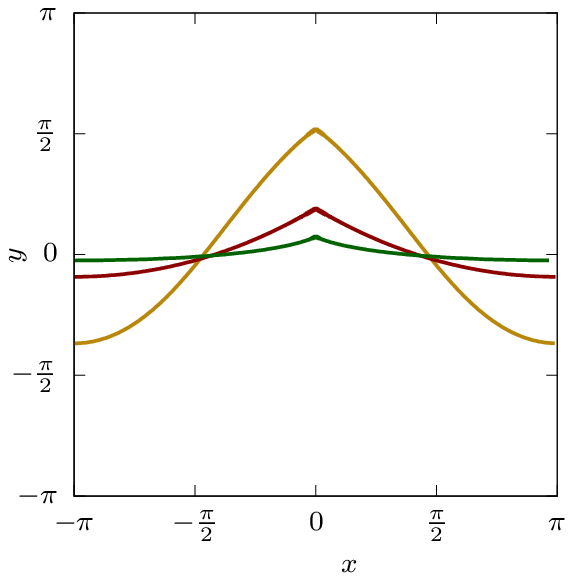}
\includegraphics[scale=1]{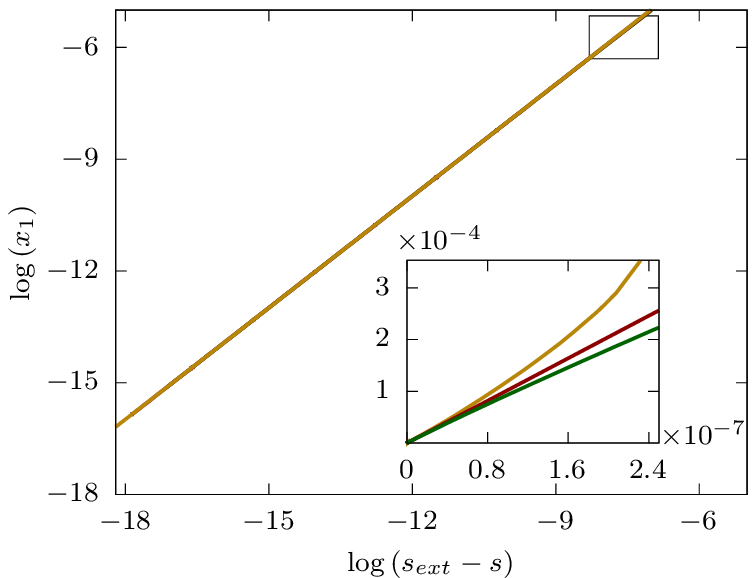}
\caption{On the left, almost extreme waves for $\omega=0$ (red), $1$ (yellow) and $-1$ (green), marked by the circles in Figure~\ref{fig:c(s)}, in the $(x,y)$ plane over the interval $x\in[-\pi,\pi]$. The mean fluid level is at $y=0$. On the right, $\log(x_1)$ vs. $\log(s_{\rm ext}-s)$ for $\omega=0$ (red), $1$ (yellow) and $-1$ (green). The inset is a close up where $s_{\rm ext}-s$ is $O(10^{-7})$.}
\label{fig6}
\end{figure}

The left panel of Figure~\ref{fig6} shows the profiles of almost extreme waves for $\omega=0$, $1$ and $-1$ in the $(x,y)$ plane over one period. We compute 
\begin{itemize}
\item[]$s=0.141063483979936080716$ ($s/s_{\rm ext}=0.99999999999999999023$) for $\omega=0$, \item[]$s=0.44310498782481126969$ ($s/s_{\rm ext}=0.99999999999999999831$) for $\omega=1$, and  \item[]$s=0.049299175088933178$ ($s/s_{\rm ext}=0.999999999999999738$) for $\omega=-1$.
\end{itemize}
The right panel of Figure~\ref{fig6} shows $x_1$ as a function of $s_{\rm ext}-s$, in the logarithmic scale, for $\omega=0$, $1$ and $-1$. Numerical evidence is clear that the thickness of the boundary layer $\propto s_{\rm ext}-s$ as $s\to s_{\rm ext}$, independently of the constant vorticity. 

\section{Conclusions}

For {\em any} constant vorticity, for the steepness sufficiently close to that of the extreme wave, we numerically find:
\begin{itemize}
\item a boundary layer where the angle the fluid surface of such {\em almost} extreme wave makes with the horizontal rises sharply from $0^\circ$ at the crest to a first local maximum, which converges monotonically to $30.3787\dots^\circ$ as $s\to s_{\rm ext}$, independently of the vorticity; the thickness of the boundary layer $\approx 102(s_{\rm ext}-s)$, independently of the vorticity; 
\item an outer region where the angle descends to $0^\circ$ at the trough for zero and negative vorticity, while it rises to a maximum $>30^\circ$ and then falls sharply to $0^\circ$ at the trough for large positive vorticity; and 
\item a transition region, where the angle oscillates about $30^\circ$, bearing resemblance to the Gibbs phenomenon; the number of oscillations increases as $s\to s_{\rm ext}$; the first local minimum angle converges monotonically to $29.9953\dots^\circ$ as $s \to s_{\rm ext}$, independently of the vorticity. 
\end{itemize}
Let $\theta_j=\theta(x_j)$ denote the $j$-th critical value of the angle in the oscillation region, where $0<x_1<x_2<\dots<x_j<\dots<\pi$. Numerical evidence suggests that $\theta_j$ converges monotonically to a limit while $|\theta_{j+1}-30^\circ|/|\theta_j-30^\circ|\to 1.22\times 10^{-2}$ as $s\to s_{\rm ext}$, for each $j$, independently of the vorticity. Also $x_j\to 0$ while $x_{j+1}/x_j \to 18.72\dots$ as $s\to s_{\rm ext}$,  for each $j$, independently of the vorticity. 

Perhaps the angle oscillations have relevance to the singularities of the conformal mapping for the Stokes wave (see \eqref{def:conformal}), where square root branch points in Riemann sheets tend to $w=0$ (corresponding to the crest) in a self-similar manner as the wave profile approaches the extreme form \citep[see, for instance,][for more discussion]{DLK2016, Lushnikov2016}. 

For zero vorticity, \cite{CG1993} solved the Nekrasov equation (see \eqref{eqn:nekrasov} efficiently to discover boundary layer and the Gibbs phenomenon near the crest of an almost extreme wave with remarkable accuracy. For nonzero constant vorticity, there is no such an integral equation, to the best of the authors' knowledge, and we instead solve the modified Babenko equation (see \eqref{eqn:babenko}) with sufficiently high accuracy.

\backsection[Funding]{This work was supported by the National Science Foundation (SD, grant number DMS-2039071), (VMH, grant number DMS-2009981).}

\backsection[Declaration of interests]{The authors report no conflict of interest.}


\backsection[Author ORCID]{S.~Dyachenko, https://orcid.org/0000-0003-1265-4055; V.~M.~Hur, https://orcid.org/0000-0003-1563-3102; D.~Silantyev, https://orcid.org/0000-0002-7271-8578}

\backsection[Author contributions]{VMH derived the theory and SD and DS performed numerical computation. All authors contributed equally to analysing data and reaching conclusions, and in writing the manuscript.}

\appendix

\section{The angle of the extreme wave}\label{appn:angle}

We give analytical evidence, similar to \citep{Stokes1880}, that for any constant vorticity, if an extreme wave has a corner at the crest then it makes a $120^\circ$ corner. 

Recall from Section~\ref{sec:prelim} that $z=x+\i y$, and we employ the notation $f=\phi+\i \psi$. Suppose for definiteness that $z_0=x_0+\i y_0$ at the crest. Suppose that
\begin{equation}\label{def:f}
f(z)=\sum_{n=0}^\infty \alpha_n (z-z_0)^n +\alpha (z-z_0)^b+o(|z-z_0|^b)\quad\text{as $z\to z_0$},
\end{equation}
where $\alpha_n,\alpha\in\mathbb{C}$ and $b\in\mathbb{R}$ is not a non-negative integer. We assume that the crest is a stagnation point, so that
\[
f_z(z_0)-\omega y_0-c=0
\]
by the second equation of \eqref{eqn:stokes}. We assume that $b>1$ and arrive at $\alpha_1=\omega y_0+c$. 

We write
\[
z-z_0=re^{i\theta},\quad \alpha_n=\rho_ne^{i\sigma_n}\quad\text{and}\quad \alpha=\rho e^{i\sigma},
\]
where $\rho_n,\rho>0$ and $\sigma_n,\sigma\in (-\pi,\pi]$. Therefore
\begin{equation}\label{eqn:phi}
\phi(r,\theta)=\sum_{n=0}^\infty \rho_nr^n\cos(n\theta+\sigma_n)+\rho r^b\cos(b\theta+\sigma)+o(r^b)
\quad\text{as $r\to0$}.
\end{equation}
Note that $\rho_1=\omega y_0+c$ and $\sigma_1=0$.

Suppose that $\theta=\theta(r)$ along the fluid surface. Suppose that 
\begin{equation}\label{eqn:theta}
\theta(r)=-\tfrac{\pi}{2}\pm\theta_0+o(1)\quad\text{as $r\to0$},
\end{equation}
where $\theta=-\frac{\pi}{2}$ bisects the angle at the crest and $2\theta_0$ measures the angle; the $+$ sign is for $r\to 0$ for $x>x_0$, and the $-$ sign for $x<x_0$. Substituting \eqref{eqn:phi} and \eqref{eqn:theta} into the third equation of \eqref{eqn:stokes}, at the leading order we gather that
\[
\tfrac12b^2\rho^22r^{2b-2}+gy_0-gr\cos\theta_0+o(r^{2b-2})=B\quad\text{as $r\to 0$}.
\]
Therefore 
\begin{equation}\label{eqn:r1}
b=\tfrac{3}{2},\quad gy_0=B\quad\text{and}\quad \tfrac12(\tfrac32)^2\rho^2-g\cos\theta_0=0.
\end{equation}
We pause to remark that $f_z(z)\propto(z-z_0)^{1/2}$ as $z\to z_0$, a square root branch point. 

To proceed, substituting \eqref{eqn:phi}, \eqref{eqn:theta} and \eqref{eqn:r1} into the second equation of \eqref{eqn:stokes}, at the order of $r^{!/2}$ we arrive at
\begin{multline*}
\tfrac32\rho r^{1/2}\cos(-\tfrac\pi4\pm\tfrac{\theta_0}{2}+\sigma)+o(r^{1/2}) \\
=(\pm\cot\theta_0+o(1))
\left(\tfrac32\rho r^{1/2}\sin(-\tfrac\pi4\pm\tfrac{\theta_0}{2}+\sigma)+o(r^{1/2})\right)\quad\text{as $r\to0$},
\end{multline*}
whence $\cos(-\frac\pi4\pm\frac32\theta_0+\sigma)=0$. Therefore $\theta_0=\frac{\pi}{3}$ and $\sigma=-\frac{3\pi}{4}$. That means, the angle at the crest is $2\theta_0=\frac{2\pi}{3}$.

\bibliographystyle{jfm}
\bibliography{stokes}

\end{document}